\documentclass[12pt]{article}

\usepackage{amsmath,amssymb,amsfonts,bbm}
\usepackage[square, comma, sort&compress]{natbib}

\newcommand{\cF}{\mathcal{F}}
\newcommand{\im}{\mathrm{i}}
\newcommand{\e}{\mathrm{e}}
\newcommand{\K}{K\"ahler{}}
\newcommand{\tr}{\mathrm{tr}}
\renewcommand{\d}{\mathrm{d}}
\newcommand{\torsion}{{\theta}}

\newcommand{\hybrid}{\topmargin -20pt
\oddsidemargin 0pt
\headheight 0pt
\headsep 0pt
\textwidth 6.25in
\textheight 9.5in
\marginparwidth .875in
\parskip 5pt plus 1pt
\jot = 1.5ex}

\hybrid

\numberwithin{equation}{section}
\numberwithin{table}{section}

\begin{document}

\begin{titlepage}

\begin{center}

\rightline{\small ZMP-HH/09-16}

\vskip 2cm

\textbf{\large Heterotic compactifications on $\mathrm{SU}(2)$-structure backgrounds}\footnote{Work supported by: DFG -- The German Science Foundation and CNCSIS -- the National University Research Council.}\\

\vskip 1.3cm

\textbf{Jan Louis$^{a,b}$, Danny Mart\'\i nez-Pedrera$^{a}$ and Andrei Micu$^c$}\\

\vskip 1.3cm

{}$^{a}${\em II. Institut f{\"u}r Theoretische Physik\\
Universit{\"a}t Hamburg\\
Luruper Chaussee 149\\
D-22761 Hamburg, Germany}\\

\vskip 0.4cm

{}$^{b}${\em Zentrum f\"ur Mathematische Physik, Universit\"at Hamburg,\\
Bundesstrasse 55, D-20146 Hamburg}

\vskip.4cm

{}$^{c}${\em Department of Theoretical Physics, \\
National Institute for Physics and Nuclear Engineering,\\
Str.~Atomistilor 407, P.O.~Box MG-6, \\
M\u{a}gurele, 077125, Romania}

\vskip 1cm 

\texttt{jan.louis@desy.de, danny@mail.desy.de, amicu@theory.nipne.ro} \\

\end{center}

\vskip 2cm

\begin{center} {\bf ABSTRACT } \end{center}
 
In this paper we study the reduction of heterotic string theory on $\mathrm{SU}(2)$-structure backgrounds.  We compute the bosonic low-energy gauged $\mathcal{N} = 2$ supergravity specified by the Killing vectors corresponding to the gauged isometries. We check that the obtained Lagrangian is consistent with the one of $\mathcal{N} = 2$ local supersymmetry. We also determine the Killing prepotentials.

\vfill

\today

\end{titlepage}

\section{Introduction}

The study of the compactification of String Theory on backgrounds with $G$-structure has received considerable attention in recent years \cite{waldram,grana}. For such backgrounds, the reduction of the structure group of the internal manifold is equivalent to the existence of one or more globally-defined internal spinors. This ensures that part of the original supersymmetry is preserved by the dimensional reduction procedure. In contrast to Calabi-Yau compactifications, these spinors need not be covariantly constant with respect to the Levi-Civita connection. Instead they are parallel with respect to a different, torsionful connection~\cite{Rocek,Hitchin,CS,salamonb}. It is in this sense that $G$-structure backgrounds represent generalizations of Calabi-Yau manifolds.

The effective theory for these compactifications turns out to be a gauged supergravity, with the aforementioned torsion playing the role of gauge charges and mass parameters. A potential is therefore generated, lifting at least part of the vacuum degeneracy typical of standard Calabi-Yau compactifications. For the heterotic string such backgrounds were first discussed in ref.~\cite{Rocek} and further considered, for example, in refs.~\cite{Dasgupta:1999ss,Lopes Cardoso:2002hd,Becker1,Becker2,Gurrieri:2007jg,Benmachiche:2008ma}.

In this paper we study compactifications of the heterotic string theory (or rather its low energy supergravity) on six-dimensional manifolds with $\mathrm{SU}(2)$-structure. Specifically, we introduce geometric twists which modify the closure relations of the harmonic one- and two-forms already present in $K3 \times T^2$. This generalization preserves eight supercharges, leading to a four-dimensional effective theory with $\mathcal{N}=2$ local supersymmetry. These manifolds have been studied previously as backgrounds of type II compactifications in refs.~\cite{waldram,Dall'Agata:2004dk,Grana:2005sn,Bovy:2005qq,ReidEdwards:2008rd,BS,TL} where they lead to $\mathcal{N}=4$ supersymmetry in four dimensions.\footnote{A more complete discussion of type II compactifications on $\mathrm{SU}(2)$-structure backgrounds will appear in ref.~\cite{LMST}.} Here we consider compactifications of the heterotic string and compute the $\mathcal{N}=2$ low energy effective action for such backgrounds.

The paper is organized as follows. In section~\ref{manisu2} we describe the features of $\mathrm{SU}(2)$-structure manifolds following refs.~\cite{waldram,Bovy:2005qq}. In order to set the stage we recall in section~\ref{k3t2} the supergravity obtained after heterotic compactification on $K3 \times T^2$ following ref.~\cite{Louis:2001uy}. In section~\ref{K3fib} we study as a first generalization the situation where a $K3$ is fibered over a torus. In this case the vector multiplet sector remains unchanged compared to compactification on $K3 \times T^2$ and modifications arise only on the hypermultiplet sector, as we show in subsection~\ref{LEEA1}. As expected, the metric of the $\sigma$-model appearing in the kinetic terms does not change, and the scalar field space coincides with that of the $K3 \times T^2$ compactification. However, certain isometries in the hypermultiplet field space which are contained in $\mathrm{SO}(4,20)$ are gauged, with the corresponding gauge bosons being the toroidal Kaluza-Klein vectors arising from reparametrizations of the torus factor. In subsection~\ref{N21} we show the consistency of the computed effective action with $\mathcal{N}=2$ gauged supergravity. In section~\ref{tau} we consider the case where some twisting is performed in the torus part as well. We can realize this case as a fibration of $K3 \times S^1$ over a circle where the monodromy is in the global symmetry group of heterotic supergravity compactified on $K3 \times S^1$. Therefore this reduction can be made sense of as a Scherk-Schwarz-type reduction~\cite{Scherk:1979zr}. In this case some isometries of the vector multiplet sector are also gauged and we compute the effective action in subsection~\ref{LEEA2}. The consistency with supergravity is checked in subsection~\ref{N22}. Section~\ref{Conc} contains our conclusions and in three appendices we give further details. In appendix~\ref{K3details} we recall the vector multiplet sector of $K3 \times T^2$ compactifications following \cite{Louis:2001uy}. In appendix~\ref{line2} we derive a formula for the line element in the space of metric deformations in terms of moduli fields that will be useful in the computation of the effective action. For completeness we compute in appendix~\ref{Pfacts} the Killing prepotentials $\mathcal{P}^x_I$ and determine their geometrical origin.

\section{Manifolds with $\mathrm{SU}(2)$ structure}
\label{manisu2}

A six-dimensional manifold $Y$ is said to have $\mathrm{SU}(2)$ structure if it admits a pair of globally-defined nowhere-vanishing $\mathrm{SO}(6)$ spinors $\eta_i$, $i=1,2$ that are linearly independent everywhere on $Y$. We will choose them to be normalized as $\bar\eta_i \eta_j = \delta_{ij}$. These spinors are the two singlets in the decomposition $\mathbf{4} \to \mathbf{2} \oplus \mathbf{1} \oplus \mathbf{1}$ of the spinor representation of $\mathrm{SO}(6)$ in representations of the reduced structure group $\mathrm{SU}(2)$. In general they are not covariantly constant with respect to the Levi-Civita connection, as is the case for Calabi-Yau manifolds such as $K3\times T^2$. Some of the properties of these manifolds were discussed in refs.~\cite{waldram,Bovy:2005qq,ReidEdwards:2008rd,BS,TL} and we summarize the results in the following.

{}From the spinor pair $\eta_i$, and using the $\mathrm{SO}(6)$ gamma-matrices $\gamma_a$, $a = 1,\ldots,6$, one can construct a triplet of self-dual two-forms $J^x$, $x=1,2,3$ and a complex one-form $v^1 + \im v^2$ as follows
\begin{equation}
\begin{aligned}
J^1_{ab} + \im J^2_{ab}  = \im \bar\eta_2 \gamma_{ab} \eta_1 \ , &\qquad J^3_{ab}  = - \tfrac\im2 (\bar\eta_1 \gamma_{ab} \eta_1 +
\bar\eta_2 \gamma_{ab} \eta_2) \ , \\ 
v^1_a + \im v^2_a & = {\bar\eta}^{\mathrm{c}}_2 \gamma_a \eta_1 \ ,
\end{aligned}
\end{equation}
where $\gamma_{ab}$ denotes the antisymmetrized product of two gamma-matrices.\footnote{Under an $\mathrm{SU}(2)$ transformation that rotates the pair of spinors $\eta_i$ into each other, the $J^x$ transform as the corresponding $\mathrm{SO}(3)$-triplet while the $v^i$ remain invariant.} The two-forms $J^x$ and the real one-forms $v^i$ characterize completely the $\mathrm{SU}(2)$-structure and are closed if and only if the spinors $\eta_i$ are covariantly constant with respect to the Levi-Civita connection. For a generic $\mathrm{SU}(2)$-structure manifold therefore, the departure from $\mathrm{SU}(2)$ holonomy (or equivalently from $K3 \times T^2$) is measured by the failure of $\d J^x$ and $\d v^i$ to vanish. The orthogonality of the $\eta_i$ together with appropriate Fierz identities implies
\begin{equation}
v^i \cdot v^{j} = \delta^{ij} \ , \qquad  J^x \wedge J^y = 2 \delta^{xy} \iota_{v^1} \iota_{v^2} \mathrm{vol}_6 \ , \qquad \iota_{v^i} J^x = 0 \ , 
\label{fierz}
\end{equation}
where $\mathrm{vol}_6$ is the volume form of $Y$ and by an abuse of notation $v^i$ denote also the vectors $v^{ia} \equiv g^{ab} v^i_a$.

Although a generic $\mathrm{SU}(2)$-structure manifold cannot be written as a product manifold like it is the case for $K3 \times T^2$, the existence of the two globally defined one-forms $v^i$ does allow us to define an almost product structure~\cite{waldram,Bovy:2005qq} 
\begin{equation}
{\Pi_a}^b = 2 v^i_a v^{ib} - \delta_a^b \ .
\label{aps}
\end{equation}
Using the first condition in \eqref{fierz} it can be easily checked that this tensor indeed satisfies ${\Pi_a}^c {\Pi_c}^b = \delta_a^b$. The tensor ${\Pi_a}^b$ splits the tangent vector over every point $p$ of $Y$ as $T_p Y = V_p \oplus W_p$, with $V_p$ and $W_p$ being two- and four-dimensional subspaces, respectively. Since ${\Pi_a}^b$ is globally defined it follows that $V = \cup_{p\in Y} V_p$ and $W = \cup_{p\in Y} W_p$ are well-defined distributions over all of $Y$. As seen from ${\Pi_a}^b v^i_b = v^i_a$, the distribution $V$ is spanned by the vector fields $v^i$. For a detailed discussion of (integrable) almost product structures see ref.~\cite{Cortes:2003zd}. In that reference a particular case of an almost product structure, namely an almost para-complex structure, is discussed.\footnote{An almost para-complex structure is an almost product structure that split the tangent space over each point into two subspaces of the same dimension.}

Integrability of the almost product structure ${\Pi_a}^b$ as encoded in the vanishing of the corresponding Nijenhuis tensor is equivalent to integrability of the distributions $V$ and $W$. This means that every neighborhood $U$ of the manifold $Y$ can be written as $U_2 \times U_4$ such that for each $p$ in $U$ we have $V_p = T_p U_2$ and $W_p = T_p U_4$, and we can introduce `separating coordinates' on every patch $U$ of $Y$ such that the metric can be given the block-diagonal structure
\begin{equation}
\d s^2 = g_{ij}(y,z)\, \d z^i \d z^j + g_{mn}(y,z)\, \d y^m \d y^n \ ,
\label{bd}
\end{equation}
where $z^i$, $i = 1,2$ are coordinates on $U_2$ and $y^m$, $m = 1, \ldots, 4$ are coordinates on $U_4$. In the following we will assume that~\eqref{aps} is integrable. The set of neighborhoods $U_2$ and $U_4$ represent foliations of the manifold $Y$, and it can happen that the leaves of these foliations are embedded submanifolds $Y_2$ and $Y_4$ of $Y$, respectively.

Together, the last condition in~\eqref{fierz} and the block-structure~\eqref{bd} for the metric forces the two-forms $J^x$ to have legs only along $U_4$. Therefore the second condition in~\eqref{fierz} becomes
\begin{equation}
J^x \wedge J^y = 2 \delta^{xy} \mathrm{vol}_4 \ ,
\end{equation}
with $\mathrm{vol}_4$ being the volume form on $U_4$. Raising an index on the two-forms $J^x$ with the metric one obtains a triplet of almost complex structures $I^x$ satisfying
\begin{equation}
I^x I^y = - \delta^{xy} \mathbbm{1} + \epsilon^{xyz} I^z \ .
\label{hyperk}
\end{equation}
Due to the spinors not being covariantly constant, these almost complex structures are in general not integrable and thus they do not form a hyperk\"ahler structure on $Y$ as they do on $K3$ or $K3\times T^2$.

In ref.~\cite{TL} the space of possible geometrical deformations of manifolds with $\mathrm{SU}(2)$ structure was discussed. If one demands the absence of massive gravitino multiplets no global one- and three-forms should exist on a four-dimensional $Y_4$. The possible deformations are then in one-to-one correspondence with the two-forms and they span the coset space
\begin{equation}
\frac{\mathrm{SO}(3,3+n)}{\mathrm{SO}(3) \times \mathrm{SO}(3+n)} \ ,
\label{su2coset}
\end{equation}
where $n$ is an integer such that the number of two-forms is $n + 6$. (For $K3$ we therefore have $n = 16$.) Furthermore, the two-forms split into three self-dual forms (which are the triplet $J^x$) and $n + 3$ anti-self-dual forms.

We therefore learn that on the six-dimensional $Y$ we have a pair of one-forms $v^i$ and $n + 6$ two-forms $\omega^A$, $A = 1, \ldots, n + 6$ at our disposal. Neither of them is necessarily closed and we shall consider the following Ansatz for their exterior derivatives~\cite{ReidEdwards:2008rd,Cvetic:2007ju}
\begin{eqnarray}
\label{algforms1} \d v^i & = & \torsion^i\, v^1 \wedge v^2 \ , \\
\label{algforms2} \d \omega^A & = & T^A_{iB}\, v^i \wedge \omega^B \ ,
\end{eqnarray}
where $\torsion^i$ and $T^A_{iB}$ are constant coefficients. In principle one could also consider adding a term proportional to $\omega^A$ to the r.h.s.~of \eqref{algforms1}, but if one insists that the almost product structure is integrable, and therefore the metric can be written as in~\eqref{bd}, such a term is ruled out. The reason for that is that $v^i$ is tangent to $U_2$, whereas as we already argued for $J^x$, the two-forms $\omega^A$ have legs only in the four-dimensional component.\footnote{Manifolds satisfying $d v^i \sim \omega^A$ and $d \omega^A = 0$ have been constructed as torus fibrations over a $K3$ base in refs.~\cite{Dasgupta:1999ss,Becker1,Goldstein:2002pg}. However, these manifolds have $\mathrm{SU}(3)$- rather than $\mathrm{SU}(2)$-structure.} 

Equation~\eqref{algforms2} implies that the two-forms $\omega^A$ are actually closed on each $U_4$. This follows since restriction to $U_4$ is achieved by setting $v^i = 0$ on the r.h.s.~of~\eqref{algforms2}, or in other words only the derivatives of $\omega^A$ in the direction of the $z^i$ (i.e.~along $U_2$) are non-trivial. This means that on each $U_4$ we have a hyperk\"ahler structure. If, as explained before, all the $U_4$ form embedded four-dimensional submanifolds $Y_4$ of $Y$, we have that each $Y_4$ must be a $K3$. As a result, the number of two-forms is no longer arbitrary but constrained by $n + 6 = 22$. We will therefore choose to focus on the cases where the $\mathrm{SU}(2)$-structure manifold is a ${K}3$ fibered over a two-dimensional space $Y_2$. We should nevertheless state that the reductions we perform here should give the same results for possible more general cases as long as eqs.~\eqref{algforms1} and \eqref{algforms2} are satisfied.

The possible values of $\torsion^i$ and $T^A_{iB}$ are restricted by the nilpotency of the $\d$-operator and by Stokes' theorem. Acting with $\d$ on~\eqref{algforms1} does not give any constraint, while acting on~\eqref{algforms2} (and using $\d^2 = 0$) yields
\begin{equation}
\torsion^i T^A_{iB} = \epsilon^{jk} T^A_{jC} T^C_{kB} \ ,
\label{cond-1}
\end{equation}
where $\epsilon_{ij} = \epsilon^{ij} = - \epsilon^{ji}$, $\epsilon^{12} = 1$.  Considering $T^A_{iB}$ as a pair of matrices $T_i \equiv (T^A_{iB})$ we can rewrite equation~\eqref{cond-1} compactly as the commutation relation
\begin{equation}
[T_1, T_2] = \torsion^i T_i \ .
\label{comm}
\end{equation}
On the other hand, Stokes' theorem implies that $\int_Y \d (v^i \wedge \omega^A \wedge \omega^B) = 0$, which after substitution of~\eqref{algforms1} and \eqref{algforms2} leads to
\begin{equation}
\epsilon^{ij} (T^A_{jC} \eta^{CB} + T^B_{jC} \eta^{CA}) = \eta^{AB} \torsion^i \ . 
\label{cond0}
\end{equation}
Here, the intersection matrix $\eta^{AB}$ is defined as
\begin{equation}
\eta^{AB} = \int_Y v^1 \wedge v^2 \wedge \omega^A \wedge \omega^B\ ,
\label{eta0}
\end{equation}
which has signature $(3,n + 3)$ as follows from the discussion of the number of self-dual and anti-self-dual two-forms on $Y$. From~\eqref{cond0} we deduce that the $T_i$ can be split as
\begin{equation}
T_i = - \tfrac12 \epsilon_{ij} \torsion^j \mathbbm{1} + {\tilde T}_i \ ,
\label{Ti}
\end{equation}
where ${\tilde T}_i$ is such that ${\tilde T}_i\eta$ is antisymmetric and thus ${\tilde T}_i$ is traceless. Since they preserve the metric $\eta^{AB}$, the ${\tilde T}_i$ are in the algebra of $\mathrm{SO}(3,n + 3)$. They also satisfy the same commutation relation~\eqref{comm} as the $T_i$, namely
\begin{equation}
[{\tilde T}_1, {\tilde T}_2] = \torsion^i {\tilde T}_i \ .
\label{comm2}
\end{equation}

In this paper we study two possible situations separately. First we consider the case $\torsion^i = 0$ (and therefore $T_i = {\tilde T}_i$) which implies
\begin{equation}
\begin{aligned}
\d v^i & = 0 \ , \\
\d \omega^A & = {\tilde T}^A_{iB}\, v^i \wedge \omega^B \ .
\end{aligned}
\label{alg1}
\end{equation}
In this case the commutation relation~\eqref{comm2} tells us that the ${\tilde T}_i$ commute and thus they form a two-dimensional Abelian subalgebra of $\mathrm{SO}(3,n+3)$. We can construct a background satisfying \eqref{alg1} if we consider $Y_2$ to be a torus as in $K3 \times T^2$ but we demand the six-dimensional manifold to be a non-trivial $K3$ fibration over this torus base. We study the reduction of heterotic supergravity on such backgrounds in section~\ref{K3fib}.

As a second case we consider a non-vanishing $\torsion^i$ in~\eqref{algforms1} but for simplicity take this time ${\tilde T}_i = 0$. As we will argue in subsection~\ref{gen}, the general case of both $\torsion^i$ and ${\tilde T}_i$ non-zero is simply a sum of these two cases. Equation~\eqref{Ti} for ${\tilde T}_i = 0$ tells us that $T^A_{iB} = - \tfrac12 \epsilon_{ij} \torsion^j \delta^A_B$ and consequently the relations~\eqref{algforms1} and \eqref{algforms2} take the form
\begin{equation}
\begin{aligned}
\d v^i & = \torsion^i v^1 \wedge v^2 \ , \\
\d \omega^A & = \tfrac12 \torsion^i \epsilon_{ij}  v^j \wedge \omega^A \ .
\end{aligned}
\label{twistedT}
\end{equation}
The first relation says that the two-dimensional component is locally a twisted torus as the one studied in ref.~\cite{Kaloper:1999yr}. It is shown in that reference that a two-dimensional twisted torus does not exist as a global manifold but here we just claim that this is a local structure in every patch $U_2$ that does not need to extend to form a whole embedded submanifold. We see that due to the second equation in~\eqref{twistedT} the $K3$ fiber is also affected by the presence of the parameter $\torsion^i$. The reduction of the heterotic string on such background will be studied in section~\ref{tau}.

In order to set the stage let us proceed by recalling the heterotic compactification on the product manifold $K3 \times T^2$.

\section{Heterotic reduction on $K3\times T^2$}
\label{k3t2}

In this section we briefly recall the derivation of the effective action for the heterotic string compactified on the product manifold $K3 \times T^2$ following ref.~\cite{Louis:2001uy}. One starts from the bosonic part of the heterotic supergravity Lagrangian in ten dimensions which is given by~\cite{Romans:1985xd}
\begin{equation}
\mathcal{L}_{10} = \e^{-\Phi} \Big( R + \d \Phi \wedge \ast \d \Phi + \tfrac12 H_3 \wedge \ast H_3 - \tfrac12 \mathrm{tr} \, F_2 \wedge \ast F_2 \Big) \ , 
\label{het10}
\end{equation}
where $\Phi$ is the ten-dimensional dilaton, $R$ is the Ricci scalar, $H_3 = \d B_2 + \ldots$ is the field strength for the NS two-form $B_2$ (the dots stand for Yang-Mills and gravitational Chern-Simons terms) and $F_2$ is the Yang-Mills field strength. The compactified theory is constrained by the consistency condition
\begin{equation}
\int_{K3} \d H = \int_{K3} (\mathrm{tr} \, R_2 \wedge R_2 - \mathrm{tr} \, F_2 \wedge F_2) = 24 - \int_{K3}\mathrm{tr} \, F_2 \wedge F_2 = 0 \ , 
\label{top}
\end{equation}
where the curvature two-form $R_2$ obeys $\int_{K3} \mathrm{tr} \, R_2 \wedge R_2 = 24$. To satisfy this constraint the gauge bundle on $K3$ has to be non-trivial in that its instanton number has to compensate the curvature contribution. This breaks part of the original non-Abelian gauge symmetry of the heterotic string. The details depend on the gauge bundle chosen, but for the purpose of this paper we do not need to be more specific and just assume that~\eqref{top} is satisfied in all cases.

The Kaluza-Klein reduction uses the following Ansatz for the metric, the NS two-form field $B_2$ and the Yang-Mills field $A^a_1$
\begin{equation}
\begin{aligned}
\d s^2 & = g_{\mu\nu} \d x^\mu \d x^\nu + g_{ij} \mathcal{E}^i \mathcal{E}^j + g_{mn} \d y^m \d y^n \ , \\ 
B_2 & = \tfrac12 B_{\mu\nu} \d x^\mu \wedge \d x^\nu + B_{i\mu} \mathcal{E}^i \wedge \d x^\mu + \tfrac12 B_{ij} \mathcal{E}^i \wedge \mathcal{E}^j + b_A \omega^A \ , \\ 
A^a_1 & = A^a_\mu \d x^\mu + A^a_i \mathcal{E}^i  \ ,
\label{ansatz}
\end{aligned}
\end{equation}
where $x^\mu$ are the coordinates of the four-dimensional space-time, $y^m$, $m = 1, \ldots, 4$ are the coordinates on $K3$ and it is convenient to define the combination $\mathcal{E}^i = \d z^i - V^i_\mu \d x^\mu$. Here the $z^i$, $i = 1,2$ are the coordinates on $T^2$ while $V^i_\mu$ are Kaluza-Klein gauge fields of $T^2$. Finally, the $\omega^A$, $A = 1,\ldots,22$ are the harmonic two-forms of $K3$.

At a generic point in the scalar field space of the compactified theory any non-Abelian gauge symmetry is broken to an Abelian subgroup $\mathrm{U}(1)^{n_{\mathrm{g}}}$. Let us identify these Abelian vector fields with the $A^a_\mu$, $a = 1, \ldots, n_{\mathrm{g}}$ in~\eqref{ansatz}. In addition there are the four KK vectors $V_\mu^i, B_{i\mu}$ and thus the effective theory contains $n_{\mathrm{v}} = 3 + n_{\mathrm{g}}$ Abelian vector multiplets (the `missing' vector being the graviphoton). The scalar superpartners in these multiplets are the $2n_{\mathrm{g}}$ scalars $A^a_i$, the four scalars contained in $g_{ij}+ B_{ij}$, the four-dimensional dilaton $\phi$ and the dual of $B_{\mu\nu}$.

The remaining scalars are assembled in $n_{\mathrm{h}}$ hypermultiplets. Twenty hypermultiplets are geometrical in that 58 out of their 80 scalars arise from the deformations of the K3 metric $g_{mn}$ and the remaining 22 from the expansion of the $B$-field denoted by $b_A$ in~\eqref{ansatz}. Additional hypermultiplets parameterize the embedding of the instanton gauge bundle inside the original ten-dimensional gauge group. The precise number and moduli space is again model-dependent, but in the following we only need to know that altogether they span a quaternionic-K\"ahler manifold of dimension $n_{\mathrm{h}}$.
  
Substituting the Ansatz~\eqref{ansatz} into the ten-dimensional Lagrangian~\eqref{het10} one derives the four-dimensional effective theory. In order to write it in the canonical $\mathcal{N} = 2$ supergravity form a number of field redefinitions have to be performed~\cite{Louis:2001uy}. Here we only give the final result but supply more details in appendix~\ref{K3details}. The bosonic Lagrangian of the compactified theory is found to be
\begin{equation}
\begin{aligned}
\mathcal{L}_4 = R + \tfrac12 I_{IJ} & F^I_{\mu\nu} F^{J\mu\nu} + \tfrac14 R_{IJ} F^I_{\mu\nu} F^J_{\rho\lambda} \epsilon^{\mu\nu\rho\lambda} - 2 G_{p\bar q}(v)\, \partial_\mu v^p \partial^\mu\bar v^{\bar q}- 2 h_{uv}(q)\, \partial_\mu q^u \partial^\mu q^v \ , 
\end{aligned}
\label{sunt}
\end{equation}
where $F^I_{\mu\nu}$, $I = 0, \ldots, n_{\mathrm{v}}$ denote the field strengths of the $n_{\mathrm{v}} + 1$ vector fields $V^i_\mu$, $B_{i\mu}$ and $A^a_\mu$ with field-dependent gauge coupling matrices $I(v)$, $R(v)$ as given in~\eqref{gaugecouplings}. The~$v^p$, $p = 1, \ldots, n_{\mathrm{v}}$ are the complex scalars in the vector multiplets which include the heterotic dilaton~$s$, the toroidal moduli~$t,u$ and the Wilson-line moduli~$n^a$. Their definition in terms of the KK-Ansatz~\eqref{ansatz} is given in~\eqref{utna}. The metric $G_{p\bar q}(v)$ is \K{} (i.e.~$G_{p\bar q} = \partial_p \bar\partial_{\bar q} K$) with \K{} potential
\begin{equation}
K = - \ln \im (s-\bar s) - \ln \tfrac14 \Big[(t-\bar t)(u-\bar u) - (n^a-{\bar n}^a)(n^a-{\bar n}^a)\Big] \ ,  
\label{Kahl}
\end{equation} 
corresponding to the coset space
\begin{equation}
\mathcal{M}_{\mathrm{v}} = \frac{\mathrm{SU}(1,1)}{\mathrm{U}(1)} \times \frac{\mathrm{SO}(2,n_{\mathrm{v}}-1)}{\mathrm{SO}(2) \times \mathrm{SO}(n_{\mathrm{v}}-1)} \ . 
\end{equation}

Finally the $q^u$, $u = 1, \ldots, 4n_{\mathrm{h}}$ in the effective action~\eqref{sunt} denote the scalars in hypermultiplets, which span a quaternionic manifold whose metric is $h_{uv}$. This metric is largely unknown due to the gauge bundle moduli. However, the 80 geometrical moduli arising as the deformations of the K3 metric and the $B$-field span the submanifold
\begin{equation}
\frac{\mathrm{SO}(4,20)}{\mathrm{SO}(4)\times\mathrm{SO}(20)}\ \subset\ \mathcal{M}_{\mathrm{h}} \ ,
\label{K3moduli}
\end{equation}
divided by a discrete symmetry group \cite{Aspinwall:1995td}. We will rederive this moduli space in the next section, where we discuss the slightly more general case of a $K3$ fibered over a torus. As we will see, the moduli space~\eqref{K3moduli} is not affected by this generalization.

\section{$K3$ fibration over torus base}
\label{K3fib}

Let us now turn to the first generalization and consider a six-dimensional manifold $Y$ constructed as a $K3$ fibered over a torus. As we discussed in section~\ref{manisu2} this corresponds to $\torsion^i=0$ in eq.~\eqref{algforms1}, or equivalently to~\eqref{alg1}. The matrices ${\tilde T}_i$ could be any two mutually commuting elements of the algebra of $\mathrm{SO}(3,19)$. Though having legs only along the $K3$ fibers, the two-forms $\omega^A (y,z)$ depend on both sets of coordinates $y^m, z^i$. They obey the differential constraint $\d \omega^A = {\tilde T}^A_{iB}\, v^i \wedge \omega^B$ but they are still harmonic on any $K3$ slice. The one-forms are $v^i = \d z^i$ and therefore satisfy $\d v^i = 0$ as required. We see that the second cohomology of the $K3$ fibers is twisted over the torus in that the basis of two-forms $\omega^A(y,z)$ changes as we go from $z$ to $z + \varepsilon$ according to
\begin{equation}
\omega^A(y, z + \varepsilon) = \omega^A(y,z) + \varepsilon^i\, {\tilde T}^A_{iB}\, \omega^B(y, z) \ .
\end{equation}
This equation can be integrated to give
\begin{equation}
\omega^A (y, z) = {(\exp{z^i {\tilde T}_i})^A}_B\, \omega^B (y, 0) \ .
\label{twist0}
\end{equation}
Once we go around the torus (choosing the identifications $z^i \sim z^i + 1$) the basis $\omega^A(y,z)$ comes back to itself up to some discrete monodromy matrices $\gamma_i$
\begin{equation}
\omega^A \to \gamma^A_{iB} \, \omega^B \ ,\qquad \gamma_i \equiv \exp{\tilde T_i} \in \Gamma(\mathbb{Z})\ ,
\label{monodromy}
\end{equation}
where for the case at hand $\Gamma(\mathbb{Z}) = \mathrm{SO}(3,19,\mathbb{Z})$ which is indeed a symmetry of the string theory.

Before we proceed let us note that as a consequence of~\eqref{monodromy} the intersection matrix $\eta^{AB}$ defined in \eqref{eta0} simplifies. Using~\eqref{twist0} and the fact that $\eta$ is an invariant metric of $\mathrm{SO}(3,19)$ we can perform the integral over the torus to arrive at
\begin{equation}
\eta^{AB} = \int_{K3} \omega^A \wedge \omega^B \ ,
\label{etadef}
\end{equation}
where we chose the normalization $\int_{T^2} \d z^i \wedge \d z^j = \epsilon^{ij}$. Thus we see that $\eta^{AB}$ reduces to the standard expression for $K3$.

\subsection{Effective action}
\label{LEEA1}

{}From the KK-Ansatz~\eqref{ansatz} and the subsequent discussion of the spectrum we infer that the twist~\eqref{twist0} does not affect the vector multiplet sector of the low energy supergravity. On the other hand, in the hypermultiplet sector it will gauge some of the isometries of the $K3$ moduli space given in~\eqref{K3moduli}. Apart from appropriate couplings to the gauge fields it will also induce a scalar potential $\mathcal{V}_\mathrm{h}$. So in the following we concentrate on the $K3$ metric moduli together with the 22 scalars $b^A$ arising from the $B$-field expansion.

In order to derive the effective action we have to consider a KK-Ansatz which slightly differs from~\eqref{ansatz} in that the metric $g_{mn}$ of the four-dimensional internal subspace now depends on the torus coordinates $z^i$. Substituting the modified Ansatz into the Ricci scalar of the ten-dimensional action~\eqref{het10} we obtain kinetic terms and a potential for the degrees of freedom in $g_{mn}$. One finds\footnote{In this paper, whenever we write an integral of a function (not a form) over any manifold $Y$ it is understood that an invariant measure of integration is used. This means that a square root of the determinant of the metric on $Y$ is included, or in other words that $\int_Y 1$ is the volume of $Y$.}~\cite{Scherk:1979zr,Maharana:1992my}
\begin{equation}
\mathcal{L}_{\mathrm{h},g} = - \tfrac14 \e^{-\phi} {\mathcal{V}_Y}^{-1} \Big( \int_{Y} g^{mp} g^{nq} \mathcal{D}_\mu g_{mn} \mathcal{D}^\mu g_{pq} + \int_{Y} g^{ij} g^{mp} g^{nq} \partial_i g_{mn} \partial_j g_{pq} \Big) \ , 
\label{reducr}
\end{equation}
where $\mathcal{D}_\mu \equiv \partial_\mu - V^i_\mu \partial_i$. Also, $\mathcal{V}_Y$ denotes the volume of $Y$. Note that on $K3\times T^2$ the metric $g_{mn}$ of $K3$ is independent of the torus coordinates $z^i$ and thus the second term in \eqref{reducr} is absent, while in the first term the $\mathcal{D}_\mu$ becomes an ordinary space-time derivative. The first term is a kinetic term for the metric degrees of freedom while the second term gives raise to a potential.

To proceed we need to rewrite the Lagrangian~\eqref{reducr} in terms of four-dimensional moduli fields.\footnote{We include this derivation explicitly since we could not find it in the literature.} In order to do so let us expand the triplet of two-forms $J^x$ defined in section~\ref{manisu2} in terms of the basis $\omega^A$ as
\begin{equation}
J^x = \e^{-\frac{\rho}2}\, \xi^x_A\, \omega^A (y,z) \ , 
\label{J}
\end{equation}
where $\xi^x_A$ are 66 real parameters and $\e^{-\rho}$ is the overall volume of the $K3$ fiber. As we discussed in section~\ref{manisu2} the $J^x$ are self dual two-forms which are singlets of the $\mathrm{SU}(2)$ structure group and satisfy
\begin{equation}
\int_{K3} J^x \wedge J^y = 2 \delta^{xy} e^{-\rho} \ .
\label{eq1}
\end{equation}
Substituting~\eqref{J} into~\eqref{eq1} and using~\eqref{etadef} we find
\begin{equation}
\eta^{AB} \xi^x_A \xi^y_B  = 2 \delta^{xy} \ .
\label{orthoxi}
\end{equation}
Thus we see that the $\xi^x_A$ are not all independent but constrained by the six equations \eqref{orthoxi}. Additionally, there is a redundancy in the possible values of $\xi^x_A$ in that an $\mathrm{SO}(3)$ rotation of the $J^x$ into each other does not take us to a new point in moduli space. Modding out this action eliminates 3 physical degrees of freedom from $\xi^x_A$. Altogether we are left with $66 - 6 - 3 = 57$ independent parameters. Adding the volume modulus $\rho$ we obtain the 58 metric moduli of $K3$ \cite{Walton:1987bu}.

Alternatively one can choose to describe the moduli in terms of the action of the Hodge star operator on the two-forms $\omega^A$. On each four-dimensional $K3$ fiber, $\ast \omega^A$ can be expanded in terms of the original $\omega^A$ basis, or in other words we have
\begin{equation}
\ast \omega^A = {M^A}_B \omega^B \ ,
\label{M0}
\end{equation}
where ${M^A}_B$ is a moduli-dependent but otherwise constant matrix. From~\eqref{etadef} one sees that
\begin{equation}
M^{AB} \equiv {M^A}_C \eta^{CB} = \int_{K3} \omega^A \wedge \ast \omega^B
\label{Msym}
\end{equation}
is symmetric. Taking the Hodge dual of~\eqref{M0} and recalling that $\ast{\ast\omega^A} = \omega^A$ one derives ${M^A}_C {M^C}_B = \delta^A_B$.  This implies that the eigenvalues of the matrix $M$ can only be $\pm 1$. Since there are three self-dual two-forms $J^x$ and nineteen anti-self-dual it follows that there must be three $+1$ and nineteen $-1$ eigenvalues. Taking the Hogde dual of \eqref{J}, using eq.~\eqref{M0} and recalling self-duality of $J^x$ we obtain $\xi^x_A {M^A}_B = \xi^x_B$, or in other words the three $\xi^x_A$ span the $(+1)$-eigenspace of ${M^A}_B$. The orthogonal subspace, i.e.~the 19-dimensional set of all $\zeta_A$ such that $\eta^{AB} \xi^x_A \zeta_B = 0$, must then be the $(-1)$-eigenspace. An operator ${M^A}_B$ that acts as the identity in the subspace spanned by the $\xi^x_A$ and as minus the identity in the orthogonal subspace, and is moreover such that $M^{AB}$ is symmetric, must necessarily be given by
\begin{equation}
\begin{aligned}
{M^A}_B & = (+) \tfrac12 \eta^{AC} \xi^x_C \xi^x_B + (-) (\delta^A_B - \tfrac12 \eta^{AC} \xi^x_C \xi^x_B) \\ 
& =  - \delta^A_B + \eta^{AC} \xi^x_C \xi^x_B \ .
\end{aligned}
\label{M}
\end{equation}
We see that ${M^A}_B$ indeed carries the information on all the metric moduli except for the volume modulus $\rho$.

Having derived the essential ingredients of the $K3$ moduli space let us return to the discussion of the six-dimensional manifolds with $\mathrm{SU}(2)$ structure and see how these moduli are affected by the specific fibration we are using. Since the $J^x$ are globally defined on the manifold we have a choice to express the twists given in \eqref{twist0} either in terms of $z$-dependent $\omega^A$ as in~\eqref{J} or equivalently by transferring the $z$-dependence to the moduli $\xi^x_A$ and $\rho$. The latter means that we can consider a basis $\omega^A(y)$ that is independent of $z^i$ and write
\begin{equation}
J^x = \e^{-\frac\rho2} \xi^x_A(z) \omega^A (y)
\end{equation}
with
\begin{equation}
\xi^x_A(z) = {(\exp{z^i {\tilde T}_i})^B}_A \xi^x_B \ .
\label{zxi}
\end{equation}
The following derivation is of course valid for both `frames', but viewing the $\omega^A$ as an honest integral basis of the second cohomology of $K3$ is often useful.\footnote{This point of view corresponds to a Scherk-Schwarz compactification where one first compactifies to six dimensions on a $K3$ and then in a subsequent step compactifies on a $T^2$ to four dimensions where the scalar fields of the six-dimensional theory have a non-trivial monodromy as one goes around the torus~\cite{Scherk:1979zr}. A similar discussion can be found in~\cite{Aharony:2008rx} for compactification of M-theory on seven-dimensional manifolds with $\mathrm{SU}(3)$ structure which can be viewed as a Calabi-Yau threefold fibered over a circle.} Note that $\rho$ does not pick up any $z$-dependence because the $\xi^x_A(z)$ just defined satisfy the normalization condition~\eqref{orthoxi}. (This will change for $\torsion^i \neq 0$ as we discuss in section~\ref{tau}.)  

In appendix~\ref{line2} we determine the line element on the space of metric deformations of the four-dimensional component $Y_4$. This can now be used to rewrite the Lagrangian given in \eqref{reducr}. We replace $\delta g$ by $\mathcal{D}_\mu g$ and $\partial_i g$ in eq.~\eqref{line3} and consequently $\mathcal{D}_\mu \xi^x_A (z)$ and $\partial_i \xi^x_A (z)$ appear in \eqref{metricf1} instead of $\delta \xi^x_A$. Similarly  $\delta\rho$ is replaced by $\partial_\mu\rho$. This leads to  
\begin{equation}
\begin{aligned}
\mathcal{L}_{\mathrm{h},g} & = - \tfrac14 \e^{-\phi} \mathcal{V}_{T^2}^{-1} \int_{T^2} \big( \partial_\mu\rho\partial^\mu\rho - 2 (\eta^{AB} - \tfrac12\, \xi^{yA} \xi^{yB})\, \mathcal{D}_\mu \xi^x_A (z) \mathcal{D}^\mu\xi^x_B (z) \big) - \mathcal{V}_\mathrm{h,g}\ , \\ 
\mathcal{V}_\mathrm{h,g} & = - \tfrac12 \e^{-\phi} \mathcal{V}_{T^2}^{-1} \int_{T^2} (\eta^{AB} - \tfrac12\, \xi^{yA} \xi^{yB})\, \partial_i \xi^x_A (z)  \partial^i \xi^x_B (z) \ ,
\end{aligned}
\label{lhg}
\end{equation}
where we substituted $\mathcal{V}_Y = \mathcal{V}_{T^2} \e^{-\rho}$, with $\mathcal{V}_{T^2}$ being the volume of the torus. In order to perform the integration over the $T^2$ we need to evaluate $\partial_i \xi^x_A(z)$. From~\eqref{zxi} we find 
\begin{equation}
\begin{aligned}
\partial_i \xi^x_A(z) & = {(\exp{z^i {\tilde T}_i})^B}_A {\tilde T}^C_{iB} \xi^x_C \ , \\
\mathcal{D}_i \xi^x_A (z) & = {(\exp{z^i {\tilde T}_i})^B}_A (\partial_\mu \xi^x_B - V^i_\mu {\tilde T}^C_{iB} \xi^x_C) \ .
\end{aligned}
\end{equation}
After substituting this derivative into eqs.~\eqref{lhg} the $z$-dependence drops out. We can intuitively see this since this dependence is all in the exponential $\exp{z^i {\tilde T}_i}$, which preserves the metric $\eta^{AB}$. The integration over the torus is now trivial and cancels the inverse torus volume factor. After performing a Weyl rescaling $g_{\mu\nu} \to \e^\phi g_{\mu\nu}$ of the four-dimensional metric we arrive at the effective four-dimensional Lagrangian
\begin{equation}
\begin{aligned}
\mathcal{L}_{\mathrm{h},g} & = - \tfrac14 \partial_\mu \rho \partial^\mu \rho + \tfrac12 (\eta^{AB} - \tfrac12 \xi^{xA} \xi^{xB}) D_\mu \xi^y_A D^\mu \xi^y_B - \mathcal{V}_{\mathrm{h},g}\\ 
& = - \tfrac14 \partial_\mu \rho \partial^\mu \rho + \tfrac18 D_\mu {M^A}_B D^\mu {M^B}_A - \mathcal{V}_{\mathrm{h},g}\ , 
\end{aligned}
\label{K}
\end{equation}
where in the second equation \eqref{M} was used. The covariant derivatives are given by
\begin{equation}
\begin{aligned}
D_\mu \xi^x_A & = \partial_\mu \xi^x_A - V^i_\mu {\tilde T}^B_{iA} \xi^x_B \ , \\ 
D_\mu M & = \partial_\mu M - V^i_\mu [M, {\tilde T}_i ] \ ,
\end{aligned}
\label{gauxim}
\end{equation}
and the potential reads
\begin{equation}
\begin{aligned}
\mathcal{V}_{\mathrm{h},g} & = - \tfrac14 \e^{\phi} g^{ij} \big( \xi^x_A {\tilde T}^A_{iB} \xi^{yB} \xi^y_C {\tilde T}^C_{jD} \xi^{xD} - 2 \xi^x_A {\tilde T}^A_{iB} {\tilde T}^B_{jC} \xi^{xC} \big) \\
& =  \tfrac14 \e^{\phi} g^{ij} \big( \tr (M {\tilde T}_i M {\tilde T}_j) - \tr ({\tilde T}_i {\tilde T}_j) \big) \\ 
& =   \tfrac18 \e^{\phi} g^{ij} \tr \big( [M, {\tilde T}_i] [M, {\tilde T}_j] \big) \ . 
\end{aligned}
\label{Vg}
\end{equation}
In the last expressions we used matrix notation with $M = ({M^A}_B)$ given in~\eqref{M} and the property $M^2 = \mathbbm{1}$. 

As a next step let us include the scalars $b_A$ arising from the $B$-field in the KK-Ansatz \eqref{ansatz}. As in~\eqref{zxi} it is useful to give them a $z$-dependence 
\begin{equation}
b_A(z) = {(\exp{z^i {\tilde T}_i})^B}_A b_B \ ,
\label{Bz}
\end{equation}
from which we compute
\begin{equation}
\partial_i b_A(z) = {(\exp{z^i {\tilde T}_i})^B}_A {\tilde T}^C_{iB} b_C \ .
\label{bi}
\end{equation}
Inserting $B_2 = b_A (z) \omega^A$ into the second term of the ten-dimensional Lagrangian~\eqref{het10} we obtain 
\begin{equation}
\mathcal{L}_{\mathrm{h},b} = - \tfrac12 \e^{-\phi} {\mathcal{V}_{T^2}}^{-1}\int_{T^2} \e^\rho  \left( \mathcal{D}_\mu b_A (z) \mathcal{D}^\mu b_B (z) + g^{ij} \partial_i b_A (z) \partial_j b_B (z) \right) \int_{K3} \omega^A \wedge \ast \omega^B \ . 
\end{equation}
Now we can insert~\eqref{bi}. The $z$-dependence drops out again and the integral over the torus is trivial. Recalling eq.~\eqref{Msym} and performing the Weyl rescaling $g_{\mu\nu} \to \e^\phi g_{\mu\nu}$ we arrive at the four-dimensional Lagrangian
\begin{equation}
\begin{aligned}
\mathcal{L}_{\mathrm{h},b} & = - \tfrac12 \e^\rho M^{AB} D_\mu b_A D^\mu b_B - \mathcal{V}_{\mathrm{h},b} \ , \\ 
\mathcal{V}_{\mathrm{h},b} & =  \tfrac12 \e^{\phi} g^{ij} \e^{\rho} b_A {\tilde T}^A_{iB} {M^{BC}} {\tilde T}^D_{jC} b_D \ , 
\end{aligned}
\label{KV}
\end{equation}
where the covariant derivative reads
\begin{equation}
D_\mu b_A = \partial_\mu b_A - V^i_\mu {\tilde T}^B_{iA} b_B \ . 
\label{gaub}
\end{equation} 

The combined Lagrangian for the metric deformation given in~\eqref{K} and for the $b$-fields given in~\eqref{KV} can be written more compactly by introducing a $24\times 24$ matrix $\mathcal{M}$ such that $\mathcal{ML}$ is symmetric and given by 
\begin{equation}
\mathcal{ML} = \begin{pmatrix} \phantom{\Big\vert} \e^\rho & \tfrac12 \e^\rho b^2 & -\e^\rho b^B \\ 
\phantom{\Big\vert} \tfrac12 \e^\rho b^2 & \e^{-\rho} + b_A M^{AB} b_B + \tfrac14 \e^\rho b^4 & - b_A M^{AB} - \tfrac12 \e^\rho b^2 b^B \\ 
\phantom{\Big\vert} -\e^\rho b^A & -M^{AB} b_B - \tfrac12 \e^\rho b^2 b^A & M^{AB} + \e^\rho b^A b^B \end{pmatrix}\ ,
\label{ML}
\end{equation}
where we abbreviated $b^2 = b_A b^A$ and defined
\begin{equation}
\mathcal{L} = \begin{pmatrix} 0 & -1 & 0 \\ -1 & 0 & 0 \\ 0 & 0 & \eta^{AB} \end{pmatrix} \ . 
\label{LL}
\end{equation}
With these conventions $\mathcal{M}$ satisfies $\mathcal{M}^2=\mathbbm{1}$ and $\mathcal{MLM}^\mathrm{T} = \mathcal{L}$ and thus $\mathcal {M}$ is an element of $\mathrm{SO}(4,20)$. Using~\eqref{ML} the complete effective Lagrangian in the hypermultiplet sector can be written as 
\begin{equation}
\begin{aligned}
\mathcal{L}_{\mathrm{h}} & = \mathcal{L}_{\mathrm{h},g} + \mathcal{L}_{\mathrm{h},b}  = \tfrac18 \mathrm{tr} \, (D_\mu \mathcal{M} D^\mu \mathcal{M}) - \mathcal{V}_{\mathrm{h}} \ , \\ 
\mathcal{V}_{\mathrm{h}} & = \mathcal{V}_{\mathrm{h},g} + \mathcal{V}_{\mathrm{h},b} = \tfrac18 \e^{\phi} g^{ij} \tr \, \big( [\mathcal{M}, \mathcal{T}_i] [\mathcal{M}, \mathcal{T}_j] \big) \ , 
\end{aligned}
\label{acttot}
\end{equation}
where
\begin{equation}
D_\mu \mathcal{M} = \partial_\mu \mathcal{M} - V^i_\mu [\mathcal{M}, \mathcal{T}_i] \ .
\end{equation}
The matrix $\mathcal{T}_i$ is defined as
\begin{equation}
\mathcal{T}_i = \begin{pmatrix} 0 & 0 & 0 \\ 0 & 0 & 0 \\ 0 & 0 & {\tilde T}_i \end{pmatrix}\ , 
\end{equation}
and thus is in the algebra of $\mathrm{SO}(4,20)$ provided that ${\tilde T}_i$ is in the algebra of $\mathrm{SO}(3,19)$.

Setting ${\tilde T}_i$ to zero corresponds to compactification on $K3 \times T^2$, and in this case the Lagrangian~\eqref{acttot} simplifies to 
\begin{equation}
\mathcal{L}_{\mathrm{h}} = \tfrac18 \mathrm{tr} \, ( \partial_\mu \mathcal{M} \partial^\mu \mathcal{M} ) \ , 
\end{equation}
which agrees with the expressions given in refs.~\cite{DLM,HLS}.

\subsection{Consistency with $\mathcal{N} = 2$ supergravity}
\label{N21}

In order to check the consistency with $\mathcal{N} = 2$ supergravity we need to compare the kinetic terms and the potential. As we already noted, for ${\tilde T}_i = 0$ the Lagrangian corresponds to compactification on $K3 \times T^2$, for which the consistency is well established. Thus we are left with checking the consistency of the covariant derivatives and the potential in~\eqref{acttot}.

For the case at hand no vector multiplets are charged and therefore the $\mathcal{N} = 2$ supergravity potential reduces to the form~\cite{Andrianopoli:1996cm}\footnote{This expression is twice the one in that reference because there the Lagrangian is normalized as $\mathcal{L}_4 = \tfrac12 R + \dotsb$.}
\begin{equation}
\mathcal{V}_\mathrm{SUGRA} = 8 \e^{K} X^I {\bar X}^J h_{uv} k^u_I k^v_J - \big[ (I^{-1})^{IJ} + 8 \e^{K} X^I {\bar X}^J \big] \mathcal{P}^x_I \mathcal{P}^x_J \ , 
\label{Vsugra}
\end{equation}
where $k^u_I$ are the Killing vectors and $\mathcal{P}^x_I$ are the corresponding Killing prepotentials defined in appendix~\ref{Pfacts}. The Killing vectors appear in the covariant derivatives of the hyper-scalars $q^u$ according to
\begin{equation}
D_\mu q^u = \partial_\mu q^u - k^u_I {\cal A}^I_\mu\ ,\qquad I = 0, \ldots, n_{\mathrm{v}}\ ,
\label{kdef}
\end{equation}
where ${\cal A}_\mu^I$ collectively denotes all vectors fields, i.e.~${\cal A}_\mu^I = (V^i_\mu, B_{i\mu}, A^a_\mu)$. The  $X^I(z)$ are related to the complex scalars of the vector multiplets as given in \eqref{Xvrel} and $I^{IJ}$ is defined in~\eqref{gaugecouplings}. Comparing \eqref{kdef} with the covariant derivatives computed in eqs.~\eqref{gauxim} and~\eqref{gaub} we conclude that 
\begin{equation}
k^\rho_{V^i} = 0 \ , \qquad k^{\xi^x_A}_{V^i} = {\tilde T}^B_{iA} \xi^x_B \ , \qquad k^{b_A}_{V^i} = {\tilde T}^B_{iA} b_B \ . 
\label{Killingv0}
\end{equation}

Note that all scalar fields are only charged with respect to $V^i_\mu$ and as a consequence the Killing vectors are non-trivial only in this direction. As shown in appendix~\ref{Pfacts} this implies that the only non-zero Killing prepotentials are $\mathcal{P}^x_{V^i}$. From this fact together with \eqref{gaugecouplings}, \eqref{utna} and \eqref{Xvrel} one shows that the negative term in the potential~\eqref{Vsugra} vanishes. With the help of eqs.~\eqref{utna} and \eqref{Xvrel} one also shows that $4 \e^\mathcal{K} X^I {\bar X}^J k^u_I k^v_J = \e^{\phi} g^{ij} k^u_{V^i} k^v_{V^j}$. Finally, the metric $h_{uv}$ can be read off from~\eqref{K} and~\eqref{KV} or equivalently from~\eqref{acttot} and is given by
\begin{equation}
h_{\rho\rho} = \tfrac18 \ , \qquad h_{\xi^x_A\xi^y_B} = - \tfrac14 (\eta^{AB} - \tfrac12 \xi^{zA} \xi^{zB}) \delta^{xy} \ , \qquad h_{b_A b_B} = \tfrac14 \e^\rho M^{AB} \ .
\label{metricc}
\end{equation}
Putting all this together we obtain
\begin{equation}
\mathcal{V}_\mathrm{SUGRA} = 2 \e^{\phi} g^{ij} h_{uv} k^u_i k^v_j = \e^{\phi} g^{ij} \Big[ \tfrac18 \tr \, \big( [M, {\tilde T}_i] [M, {\tilde T}_j] \big) + \tfrac12 \e^\rho b_A {\tilde T}^A_{iB} M^{BC} {\tilde T}^D_{jC} b_D \Big] \ . 
\end{equation}
This expression is in complete agreement with $\mathcal{V}_\mathrm{h}$ in eq.~\eqref{acttot}.

\section{$\mathrm{SU}(2)$-structure compactifications with $\torsion^i \neq 0$.}
\label{tau}

In this section we consider the case where $\torsion^i\neq 0$ and $\tilde T_i=0$, that is we impose the differential relations $\d v^i = \torsion^i v^1 \wedge v^2$ and $\d \omega^A = \tfrac12 \torsion^i \epsilon_{ij} v^j \wedge \omega^A$ as given in eqs.~\eqref{twistedT}. Allowing additionally for ${\tilde T}_i \neq 0$ simply combines the results of the former section to what will be found here. We discuss this case briefly at the end of this section.

Without loss of generality we can assume that $\torsion^2 = 0$ so that only the first component $\torsion^1 \equiv \torsion$ is non-zero. This can always be achieved by an $\mathrm{SO}(2)$ rotation of the pair of one-forms $v^i$. Thus we have $\d v^1 = \torsion v^1 \wedge v^2$, $\d v^2 = 0$ together with $\d \omega^A = \tfrac12 \torsion v^2 \wedge \omega^A$, which are satisfied by
\begin{equation}
v^1 = \e^{-\torsion z^2} \d z^1 \ , \qquad v^2 = \d z^2 \ , \qquad \omega^A(z) = \e^{\tfrac12 \torsion z^2} \omega^A \ . 
\label{vvom}
\end{equation}

We could construct such background by considering $K3 \times S^1$ fibered over a second circle parametrized by $z^2 \sim z^2 + 1$. The circle in the fiber has $z^1$ as coordinate and together with the base circle they have locally the structure of a two-dimensional twisted torus as considered in~\cite{Kaloper:1999yr}. As already mentioned though, a two-dimensional twisted torus does not exist as a global manifold.

On the $K3$ part of the fiber we can perform an expansion similar to~\eqref{J} and transfer the $z$-dependence of $\omega^A(z)$ to the moduli $\rho$ and $\xi^x_A$. In view of the third equation in~\eqref{vvom} we conclude that we have to set
\begin{equation}
\rho(z) = \rho - \torsion z^2 \ ,
\label{rho2}
\end{equation}
while the $\xi^x_A$ remain independent of $z^i$. This simply reflects the fact that~\eqref{vvom} demands a rescaling of the $J^x$. From the term $b_A \omega^A$ in the expansion of the NS two-form we conclude that the $b$-fields must be given a $z$-dependence
\begin{equation}
b_A(z) = \e^{\tfrac12 \torsion z^2} b_A \ .
\label{bz}
\end{equation}

The question now arises of how to patch the fibers after going once around the base circle $z^2 \to z^2 + 1$, i.e.~how to make sense of the monodromy. We will see that this identification is possible if we consider the fact that heterotic supergravity compactified to six-dimensions on $K3$ has indeed a global $\mathrm{SO}(4,20)$ symmetry (which  gets broken to a discrete subgroup thereof in the full heterotic string theory). Leaving aside for the moment this issue let us start with the derivation of the effective action.

\subsection{Effective action}
\label{LEEA2}

The difference compared to the situation discussed in the previous section \ref{K3fib} are the twisted differential relations of the one-forms $\d v^1 = \torsion v^1 \wedge v^2, \d v^2=0$. They have the effect that also isometries of the manifold spanned by scalars in vector multiplets are gauged. In the KK-reduction we can largely follow the analysis of ref.~\cite{Kaloper:1999yr} where the heterotic string compactified on twisted tori was considered. Without repeating the derivation in detail here let us state that the covariant derivatives which follow from ref.~\cite{Kaloper:1999yr} are
\begin{equation}
\begin{aligned}
D_\mu g_{ij} & = \partial_\mu g_{ij} + \torsion g_{1i} \epsilon_{jk} V^k_\mu + \torsion g_{1j} \epsilon_{ik} V^k_\mu \ , \\ 
D_\mu B_{12} & = \partial_\mu B_{12} - \torsion B_{1\mu} + \torsion B_{12} V^2_\mu \ , \\ 
D_\mu A^a_i & = \partial_\mu A^a_i + \torsion A^a_1 \epsilon_{ij} V^j_\mu \ , 
\end{aligned}
\label{gaugs1}
\end{equation}
while the axion-dilaton $s$ remains neutral. If we express the covariant derivatives \eqref{gaugs1} in the complex variables $u$, $t$ and $n^a$ defined in \eqref{utna} we obtain
\begin{equation}
\begin{aligned}
D_\mu u & = \partial_\mu u - \torsion (u V^2_\mu + V^1_\mu) \ , \\
D_\mu t & = \partial_\mu t + \torsion (t V^2_\mu - B_{1\mu}) \ ,
\end{aligned}
\label{covdevvec2}
\end{equation}
while the fields $n^a$ remain neutral. 

Furthermore, a potential $\mathcal{V}_\mathrm{v}$ is generated and given by
\begin{equation}
\mathcal{V}_{\mathrm{v}} = {} \e^{\phi} \vert g \vert^{-1} (g_{11} + \tfrac12 A^a_1 A^a_1) \torsion^2 = {} - \torsion^2  \e^K \frac{(u-\bar u) (t-\bar t)}{(u-\bar u) (t-\bar t) - (n^a - {\bar n}^a)^2} \ .
\label{vvector}
\end{equation}
Here $g_{ij}$ is the metric of the two-dimensional component of $Y$ and $\vert g\vert \equiv \det g_{ij}$. The final expression is written in terms of the complex vector moduli $u$, $t$ and $n_a$ and the four-dimensional dilaton $\phi$ as defined in appendix~\ref{K3details}. This result is consistent with the potential derived in ref.~\cite{Kaloper:1999yr} if applied to a twisted two-torus.

Since $\torsion$ also appears in the differential relations for $\omega^A$ in \eqref{twistedT} the hypermultiplet sector is similarly affected. Repeating the analysis of the previous section one finds that only the volume modulus $\rho$ and the $b_A$ fields acquire a charge. Considering the action of $\mathcal{D}_\mu = \partial_\mu - V^i_\mu \partial_i$ on eqs.~\eqref{rho2} and \eqref{bz} we find the following covariant derivatives for $\rho$ and $b_A$
\begin{equation}
\begin{aligned}
D_\mu \rho & = \partial_\mu \rho + \torsion V^2_\mu \ , \\
D_\mu b_A & = \partial_\mu b_A - \tfrac12 \torsion V^2_\mu b_A \ .
\end{aligned}
\label{gaug2}
\end{equation}
As already seen the moduli comprised in $\xi^x_A$ or equivalently ${M^A}_B$ remain neutral and we can encode all covariant derivatives in the expression
\begin{equation}
D_\mu \mathcal{M} = \partial_\mu \mathcal{M} - V^2_\mu [\mathcal{M}, \mathcal{T}] \ , 
\label{covM}
\end{equation}
with $\mathcal{M}$ given in eq.~\eqref{ML} and the matrix $\mathcal{T}$ in the algebra of $\mathrm{SO}(4,20)$ defined by
\begin{equation}
\mathcal{T} = \begin{pmatrix} - \tfrac12 \torsion & 0 & 0 \\ 0 & \tfrac12 \torsion & 0 \\ 0 & 0 & 0 \end{pmatrix} \ . 
\label{Ti2}
\end{equation}

We pause here to annotate the following. Since $\mathrm{SO}(4,20)$ is a global symmetry of heterotic supergravity compactified to six-dimensions on $K3$ we can indeed make sense of this background as a Scherk-Schwarz type of reduction in which the $K3$ sigma-model moduli organized in the matrix $\mathcal{M}$ pick up a monodromy $\e^{\mathcal{T}}$ in further compactifications on circles. Unfortunately there is no non-vanishing value of $\torsion$ such that $\e^\mathcal{T}$ is in $\mathrm{SO}(4,20,\mathbb{Z})$, so the lifting to string theory is unclear. 

A potential $\mathcal{V}_\mathrm{h}$ in the hypermultiplet sector is also generated. It has two contributions, one from the reduction of the ten-dimensional Ricci scalar due to the dependence of the four-dimensional volume modulus $\rho$ on the two-dimensional local coordinates, and a second one involving the fields $b_A$ arising from the kinetic term for the ten-dimensional $B$-field. Putting them together we obtain the following expression for the potential,
\begin{equation}
\mathcal{V}_{\mathrm{h}} = \tfrac14 \e^{-\phi} \vert g \vert^{-1} g_{11} \torsion^2 ( 1 + \tfrac12 \e^\rho b^\mathrm{T} M b ) = \tfrac14 \torsion^2 \e^{K} ( 1 + \tfrac12 \e^\rho b^\mathrm{T} M b ) \ . 
\label{vhyper}
\end{equation}
It can be checked that this potential together with the kinetic terms can be cast in the following form for the Lagrangian
\begin{equation}
\mathcal{L}_{\mathrm{h}} = - \tfrac18 \mathrm{tr} (D_\mu \mathcal{M} D^\mu \mathcal{M}) + \tfrac18 \e^{K} \tr \big( [\mathcal{M}, \mathcal{T}] [\mathcal{M}, \mathcal{T}] \big) \ , 
\end{equation}
where the covariant derivative is given in eq.~\eqref{covM}, the matrix $\mathcal{T}$ in eq.~\eqref{Ti2} and $K$ is the \K{} potential~\eqref{Kahl} for the scalar manifold corresponding to the vector-multiplet sector.

\subsection{Consistency with $\mathcal{N} = 2$ supergravity}
\label{N22}

We can check the consistency with $\mathcal{N} = 2$ supergravity in much the same way as we did in section~\ref{N21}. Now we have contributions to the potential of the theory coming from both sectors, $\mathcal{V}_\mathrm{v}$ and $\mathcal{V}_\mathrm{h}$. The supergravity potential for the hypermultiplets was already given in~\eqref{Vsugra} and the fact that the negative term vanishes remains valid also in this case. For the vector multiplets the scalar potential is positive definite and again proportional to the Killing vectors. Together they read~\cite{Andrianopoli:1996cm} 
\begin{equation}
\mathcal{V}_\mathrm{SUGRA} = 2 \e^K X^{\bar I} X^{J} (G_{p\bar q} k^p_I k^{\bar q}_J + 4 h_{uv} k^u_I k^v_J) \ , 
\label{Vsugrav}
\end{equation}
where $k^p_I$ are the Killing vectors for the vector multiplets and $G_{p\bar q}$ the K\"ahler metric derived from the K\"ahler potential~\eqref{Kahl}. The generic covariant derivatives for the hyper-scalars are defined in~\eqref{kdef} and so for the vector multiplet scalars one defines analogously  
\begin{equation}
D_\mu v^p = \partial_\mu v^p - k^p_I {\cal A}^I_\mu\ , \qquad p=1,\ldots, n_{\mathrm{v}}\ ,
\label{kidef}
\end{equation}
where  $v^p$ collectively denotes all scalars, i.e.~$v^p = (s, u, t, n^a)$. Comparing \eqref{kdef} and \eqref{kidef} with~\eqref{covdevvec2} and \eqref{gaug2} we arrive at 
\begin{equation}
\begin{aligned}
k^{p}_{V^2} & = (0, \torsion u, - \torsion t, 0) \ , \quad k^p_{V^1} = (0, \torsion, 0, 0) \ , \quad k^p_{B_1} = (0, 0, \torsion, 0) \ , \\
k^\rho_{V^2} & = - \torsion \ , \quad k^{b_A}_{V^2} = \tfrac12 \torsion b_A \ , 
\end{aligned}
\label{Killingv}
\end{equation}
while all other Killing vectors vanish. Inserting~\eqref{Killingv}, \eqref{metricc}  and $G_{p\bar q}$ obtained as the second derivative of the \K{} potential given in \eqref{Kahl} into \eqref{Vsugrav} we can check straightforwardly that $\mathcal{V}_\mathrm{SUGRA}$ coincides with $\mathcal{V}_\mathrm{v} + \mathcal{V}_\mathrm{h}$ as given in eqs.~\eqref{vvector} and \eqref{vhyper}.

Moreover, we can compute the Killing prepotential $\mathcal{P}_I$ for the vector multiplet sector. This prepotential is real and must satisfy the equation
\begin{equation}
k^p_I = \im G^{p\bar q} \partial_{\bar q} \mathcal{P}_I \ .
\label{prepoteq}
\end{equation}
It can be checked that if we substitute in the r.h.s.~of eq.~\eqref{prepoteq} the expressions
\begin{equation}
\begin{aligned}
\mathcal{P}_{V^1} & = \im \torsion \frac{t - \bar t}{(u - \bar u)(t - \bar t) - (n^a - {\bar n}^a)^2} \ , \\
\mathcal{P}_{V^2} & = \im \torsion \frac{\bar u t - u \bar t}{(u - \bar u)(t - \bar t) - (n^a - {\bar n}^a)^2} \ , \\
\mathcal{P}_{B_1} & = \im \torsion \frac{u - \bar u}{(u - \bar u)(t - \bar t) - (n^a - {\bar n}^a)^2} \ ,
\end{aligned}
\end{equation}
we indeed obtain the Killing vectors given in the first line of~\eqref{Killingv}.

\subsection{General case with both $\torsion^i$ and ${\tilde T}_i$ non-zero}
\label{gen} 

Allowing for non-vanishing ${\tilde T}_i$ amounts to modifying the third equation in~\eqref{vvom} in such a way that it satisfies
\begin{equation}
\begin{aligned}
\d \omega^A & = \tfrac12 \torsion v^2 \wedge \omega^A + {\tilde T}^A_{iB} v^i \wedge \omega^B \\ 
& = \tfrac12 \torsion \d z^2 \wedge \omega^A + {\tilde T}^A_{2B} \d z^2 \wedge \omega^B + {\tilde T}^A_{1B} \e^{-\torsion z^2} \d z^1 \wedge \omega^B \ , 
\end{aligned}
\label{fin}
\end{equation}
where the one-forms $v^i$ are given in~\eqref{vvom} but now we additionally allow for matrices ${\tilde T}_i$ having the non-zero commutator
\begin{equation}
[{\tilde T}_1, {\tilde T}_2] = \torsion {\tilde T}_1
\label{tt}
\end{equation}
as follows from~\eqref{comm2}. We can check that~\eqref{fin} is satisfied if we set
\begin{equation}
\omega^A(z) = \e^{\frac12 \torsion z^2} {(\exp{z^2 {\tilde T}_2})^A}_B {(\exp{z^1 {\tilde T}_1})^B}_C \omega^C \ . 
\label{aa}
\end{equation}
The first two terms in the last equality of~\eqref{fin} are easily seen to arise from~\eqref{aa}. The third term arises as well if we compute 
\begin{equation}
\begin{aligned}
\partial_1 \omega^A(z) & = \e^{\frac12 \torsion z^2} {(\exp{z^2 {\tilde T}_2})^A}_B {\tilde T}^B_{1C} {(\exp{z^1 {\tilde T}_1})^C}_D \omega^D \\ 
& = \e^{\frac12 \torsion z^2} \e^{-\torsion z^2} {\tilde T}^A_{1B} {(\exp{z^2 {\tilde T}_2})^B}_C {(\exp{z^1 {\tilde T}_1})^C}_D \omega^D \\ 
& = \e^{-\torsion z^2} {\tilde T}^A_{1B} \omega^B(z) \ ,
\end{aligned}
\end{equation}
where we have used
\begin{equation}
\exp(z^2 {\tilde T}_2) \, {\tilde T}_1 \, \exp(- z^2 {\tilde T}_2) = \e^{- \torsion z^2} {\tilde T}_1 \ ,
\end{equation}
as follows from the commutation relation~\eqref{tt}.

In the frame where the $\omega^A$ are $z$-independent we derive from eq.~\eqref{aa} the $z$-dependent moduli to be 
\begin{equation}
\rho(z) = \rho - \torsion z^2 \ , \qquad \xi^x_A(z) = {(\exp{z^2 {\tilde T}_2})^C}_B {(\exp{z^1 {\tilde T}_1})^B}_A \xi^x_C \ . 
\label{split}
\end{equation}
This is the proper splitting of the $z$-dependence between $\rho$ and $\xi^x_A$ because it satisfies the orthonormality constraint~\eqref{orthoxi}. Comparing~\eqref{split} with~\eqref{zxi} and \eqref{rho2} we see that for the general case one simply has a `sum' of the gaugings obtained for $\torsion = 0$ in section~\ref{K3fib} and ${\tilde T}_i = 0$ in this section.

We can now combine the Killing vectors~\eqref{Killingv0} and~\eqref{Killingv} to find the gauge algebra for the general case. The generators of the gauge algebra are constructed from the Killing vectors as
\begin{equation}
k_I = k^p_I \partial_p + k^u_I \partial_u \ ,
\end{equation}
where $k^p_I$ and $k^u_I$ are the Killing vectors of the gauged isometries of the scalar spaces of vector multiplets and hypermultiplets, respectively. The derivatives are with respect to the vector multiplet scalars $v^p = (s, u, t, n^a)$ and the hyper-scalars $q^u = (\rho, \xi^x_A, b_A)$. Substituting eqs.~\eqref{Killingv0} and~\eqref{Killingv} we obtain
\begin{equation}
\begin{aligned}
k_{V^1} & = \xi^x_A {\tilde T}^A_{1B} \frac{\partial}{\partial \xi^x_B} + b_A {\tilde T}^A_{1B} \frac{\partial}{\partial b_B} + \torsion \frac{\partial}{\partial u} \ , \\
k_{V^2} & = \xi^x_A {\tilde T}^A_{2B} \frac{\partial}{\partial \xi^x_B} + b_A {\tilde T}^A_{2B} \frac{\partial}{\partial b_B} - \torsion \frac{\partial}{\partial \rho} + \tfrac12 \torsion b_A \frac{\partial}{\partial b_A} + \torsion \frac{\partial}{\partial u} - \torsion t \frac{\partial}{\partial t} \ , \\
k_{B_1} & = \torsion \frac{\partial}{\partial t} \ ,
\end{aligned}
\end{equation}
as the only non-zero components. From these expressions and recalling~\eqref{comm2} we can compute the commutation relations
\begin{equation}
[k_{V^1}, k_{V^2}] = \torsion k_{V^1} \ , \quad [k_{V^1}, k_{B_1}] = 0 \ , \quad [k_{V^2}, k_{B_1}] = \torsion k_{B_1} \ .
\end{equation}
This is a solvable and therefore not semisimple algebra. As expected, when $\torsion = 0$ it turns into an Abelian algebra.

\section{Conclusions}
\label{Conc}

In this paper we have derived the four-dimensional gauged $\mathcal{N} = 2$ supergravities arising from the reduction of heterotic string theory on backgrounds with $\mathrm{SU}(2)$ structure. The backgrounds studied have been obtained by twisting the cohomology of $K3 \times T^2$, i.e.~by writing the action of the $\d$-operator on the set of one- and two-forms as linear combinations of exterior products of the forms themselves.

Specifically we have studied two independent classes of such twists. In the first case we have considered twisting the $K3$ harmonic forms $\omega^A$ by elements of the SO(3,19) symmetry group which rotates these forms among themselves as going around the $T^2$ base. This leads to gaugings in the hypermultiplet moduli space which is spanned by the $K3$ moduli.

The second case we have analyzed can be understood as compactification to five dimensions on $K3 \times S^1$ followed by a Scherk-Schwarz compactification on another $S^1$ to four dimensions. In this case the full $K3 \times S^1$ is twisted as going around the second $S^1$ and as a result both isometries in the hyper and the vector moduli spaces are gauged. We have checked in both cases the resulting action against the general $\mathcal{N}=2$ gauged supergravity in four dimensions. 

As mentioned in section \ref{tau}, in the second case the full embedding into string theory is problematic as for this, the twist after going around the entire $S^1$ has to be in the (discrete) U-duality group of the compactification, which we have seen that does not happen. Nevertheless, from the supergravity point of view the compactification to four dimensions on the final $S^1$ is fully consistent as also shown by the four-dimensional result which is in agreement with $\mathcal{N}=2$ gauged supergravity.

It is worth mentioning that the gauging in the vector multiplet sector obtained in the second case is the same as one specific case of~\cite{Aharony:2008rx}. The twisting on the M-theory side of~\cite{Aharony:2008rx} precisely corresponds to that obtained by an element~\eqref{Ti2}, which we have seen that can not exponentiate to an element of the integer U-duality group of string theory. Therefore on both sides the embedding into string/M-theory is problematic and thus the fact that the gaugings are the same may be purely accidental and without any meaning in the context of string dualities. This final point seems to be confirmed by the fact that in the case presented in this paper the vector multiplet gaugings requires a gauging in the hypermultiplet sector, while this does not seem to be the case in~\cite{Aharony:2008rx} where only vector multiplet gaugings were obtained.

\vskip 1cm

\subsection*{Acknowledgements}

The work of JL and DM was supported by the Deutsche Forschungsgemeinschaft (DFG) in the SFB 676 ``Particles, Strings and the Early Universe''. The work of AM was supported by the National University Research Council (CNCSIS/UEFISCSU) through the reintegration grant 3/3.11.2008 and program "Idei", contract number 464/15.01.2009. 

We have greatly benefited from conversations with R.~Reid-Edwards, B.~Spanjaard, H.~Triendl and T.~Danckaert.

\clearpage

\appendix

\noindent
\textbf{\Large Appendix}

\section{Vector multiplet sector in heterotic $K3\times T^2$ compactifications}
\label{K3details} 

In this appendix we provide more details of the vector multiplet sector in heterotic compactifications on $K3\times T^2$ following~\cite{Louis:2001uy}.

In order to derive the Lagrangian \eqref{sunt} one substitutes~\eqref{ansatz} into the ten-dimensional Lagrangian \eqref{het10}. This yields
\begin{equation}
\begin{aligned}
\mathcal{L}_4 = R + \tfrac12 I_{IJ} & F^I_{\mu\nu} F^{J,\mu\nu} + \tfrac14 R_{IJ} F^I_{\mu\nu} F^J_{\rho\lambda} \epsilon^{\mu\nu\rho\lambda} \\ 
& + \frac{2 \partial_\mu s \partial^\mu \bar s}{(s-\bar s)^2} + \tfrac1{8} \partial_\mu M_{IJ} \partial^\mu M^{IJ} - 2 h_{uv} \partial_\mu q^u \partial^\mu q^v \ . 
\end{aligned}
\label{suntA}
\end{equation}
Here $M^{IJ}$ is an $\mathrm{SO}(2,n_{\mathrm{v}}-1)$ matrix  of the form
\begin{equation}
M^{IJ} = \begin{pmatrix} G^{-1} & - G^{-1} C & - G^{-1} A \\ 
- C^\mathrm{T} G^{-1} & G + A A^\mathrm{T} + C^\mathrm{T} G^{-1} C & A + C^\mathrm{T} G^{-1} A \\
- A^\mathrm{T} G^{-1} & A^\mathrm{T} + A^\mathrm{T} G^{-1} C & \mathbbm{1}_{n_{\mathrm{v}}-3} + A^\mathrm{T} G^{-1} A \end{pmatrix} \ , 
\end{equation}
where $G = (g_{ij})$, $B = (B_{ij})$, $A = (A^a_i)$ and we abbreviated $C = B + \tfrac12 A A^\mathrm{T}$. The inverse $M_{IJ}$ is obtained by lowering the indices with the metric of $\mathrm{SO}(2,n_{\mathrm{v}}-1)$
\begin{equation}
L_{IJ} = \begin{pmatrix} 0 & \mathbbm{1}_2 & 0 \\ \mathbbm{1}_2 & 0 & 0 \\ 0 & 0 & \mathbbm{1}_{n_{\mathrm{v}}-3} \end{pmatrix} \ , 
\label{L}
\end{equation}
which is left invariant by $M$ or in other words $M L M = L$ holds. The one additional scalar field~$s$ in~\eqref{suntA} also is a member of a vector multiplet. It is defined as the combination
\begin{equation}
s = \frac a2 - \frac\im2 \e^{-\phi} \ ,
\end{equation}
where $a$ is the axion dual to $B_{\mu\nu}$ and $\phi$ is the four-dimensional dilaton defined as $e^{-\phi}=e^{-\Phi} \mathrm{vol}_6$. The gauge couplings in \eqref{suntA} are found to be 
\begin{equation}
I_{IJ} = \frac{(s-\bar s)}{2\im} M_{IJ} \ , \qquad R_{IJ} = - \frac{(s+\bar s)}2 L_{IJ} \ .
\label{gaugecouplings}
\end{equation}
The scalar fields $g_{ij}$, $B_{ij}$ and the $A^a_i$ can be traded for the complex \K{} coordinates $u, t$ and $n^a$ by the field redefinition~\cite{Louis:2001uy}
\begin{equation}
\begin{aligned}
g_{11} & = \frac{2\im}{u-\bar u}\sqrt{\vert g\vert} \ , \qquad g_{12} = \im \frac{u+\bar u}{u-\bar u} \sqrt{\vert g\vert}\ , \\ 
\sqrt{\vert g\vert} & = - \frac\im2 \Big[ (t-\bar t) - \frac{(n^a-{\bar n}^a) (n^a-{\bar n}^a)}{u - \bar u} \Big] \ , \\ 
B_{12} & = \frac12 \Big[ (t+\bar t) - \frac{(n^a+{\bar n}^a)(n^a-{\bar n}^a)}{u - \bar u} \Big] \ , \\
A^a_1 & = \sqrt2 \, \frac{n^a-{\bar n}^a}{u-\bar u}\ , \qquad A^a_2 = \sqrt2 \, \frac{\bar u n^a-u{\bar n}^a}{u-\bar u} \ . 
\end{aligned}
\label{utna}
\end{equation}
Inserting \eqref{utna} into \eqref{suntA} one arrives at the terms involving the vector multiplets of the Lagrangian given in~\eqref{sunt}.

The Lagrangian given in \eqref{sunt} is not of the standard supergravity form but can only be obtained from it after an appropriate symplectic rotation~\cite{Louis:2001uy}. In $\mathcal{N} = 2$ supergravity the K\"ahler potential $K$ is determined in terms of a holomorphic prepotential $\cF$ according to~\cite{dWvP}
\begin{equation}
K = - \ln \Big[ \im \bar{X}^{I} (\bar v) \cF_{I}(X)  - \im X^{I} (v) \bar{\cF}_{I}(\bar{X}) \Big] \ . 
\label{Kspecial}
\end{equation}
The $X^{I}$, $I = 0,\ldots, n_v$ are $(n_v+1)$ holomorphic functions of the scalars $v^p$, and $\cF_{I}$ abbreviates the derivative, i.e.~$\cF_{I}\equiv \frac{\partial \cF(X)}{\partial X^{I}}$. Furthermore $\cF(X)$ is a homogeneous function of degree $2$ in $X^{I}$, i.e.\ $X^{I} \cF_{I}=2 \cF$.

For the case at hand the $X^I(v)$ are related to the complex fields $u, t$ and $n^a$ as follows,
\begin{equation}
X^0 = \tfrac12 t\ , \quad  X^1 = \tfrac12(ut - n^a n^a)\ , \quad X^2 = - \tfrac12 u\ , \quad X^3 = \tfrac12\ , \quad  X^a = \tfrac1{\sqrt2} n^a\ , 
\label{Xvrel}
\end{equation}
which is a symplectic rotation from a more standard basis. (See~\cite{Louis:2001uy} for more details. Also note that this convention differs from the one of~\cite{Louis:2001uy} so that we can use $L_{IJ}$ as defined in~\eqref{L}.)

\section{Derivation of line element in the space of metrics}
\label{line2} 

In this appendix we give a derivation of an expression for the line element
\begin{equation}
\delta s^2 = \int_{Y_4} \! \sqrt{\vert g \vert} \, g^{mn} g^{pq} \delta g_{mp} \delta g_{nq} 
\label{line3}
\end{equation} 
in the space of metrics $g_{mn}$ in terms of the variations of the moduli $\rho$ and $\xi^x_A$ as defined in eq.~\eqref{J}, or equivalently in terms of ${M^A}_B$ as defined in eq.~\eqref{M0}. Although in the main text we mostly apply it to $Y_4 = K3$ the following derivation holds more generally. As in eq.~\eqref{J} we expand $J^x = \e^{-\frac\rho2} \xi^x_A \omega^A $ with $\eta^{AB} \xi^x_A \xi^y_B = 2 \delta^{xy}$ and $\eta^{AB}$ being the intersection matrix of the $\omega^A$ defined in eq.~\eqref{etadef}. 

Raising an index on $(J^x)_{mn}$ by means of the metric we can define a triplet of almost complex structures ${(I^x)_m}^n = (J^x)_{mp} g^{pn}$ satisfying eq.~\eqref{hyperk}, that is
\begin{equation}
{(I^1)_m}^p {(I^1)_p}^n = - \delta_m^n \ , \qquad {(I^1)_m}^p {(I^2)_p}^n = {(I^3)_m}^n \ , 
\label{hyperk2}
\end{equation}
and cyclic permutations thereof. In the following it will prove convenient to work in matrix notation and set $J^x = (J^x)_{mn}$, $I^x = {(I^x)_m}^n$ and $g = g_{mn}$. We can therefore write for example $I^x g = J^x$. If we act on the left of this equality with $I^x$ and use the first equation in~\eqref{hyperk2} we obtain $g = - I^1 J^1 = - I^2 J^2 = - I^3 J^3$. Thus the variation $\delta g$ is given by
\begin{equation}
\begin{aligned}
\delta g & = - I^1 \delta J^1 - \delta I^1 J^1 = - I^2 \delta J^2 - \delta I^2 J^2  = - I^3 \delta J^3 - \delta I^3 J^3 \ . 
\label{deltag}
\end{aligned}
\end{equation}
The variation of the second equation in~\eqref{hyperk2} yields
\begin{equation}
\delta I^3 = \delta I^1 I^2 + I^1 \delta I^2 \ .
\end{equation}
{}From this expression and making repeated use of~\eqref{hyperk2} we derive
\begin{equation}
\begin{aligned}
\delta I^3 J^3 & = I^1 (\delta I^1 J^1 - \delta I^2 J^2) g^{-1} I^1 \\
& = (\delta J^1 + I^3 \delta J^2) g^{-1} J^1 \ ,
\end{aligned}
\label{deltaint}
\end{equation}
where in the last step we used the second equality in~\eqref{deltag}. Substituting \eqref{deltaint} into the last equality of~\eqref{deltag} we arrive at
\begin{equation}
\delta g = - I^3 \delta J^3 - (\delta J^1 + I^3 \delta J^2) g^{-1} J^1 \ .
\label{master}
\end{equation}
This expresses $\delta g$ in terms of $\delta J^x$. Clearly a similar expression can be given with $I^x$ and $J^x$ cyclically permuted. The variations of the $J^x$ are all independent with the exception of the volume modulus. Using the cyclic symmetry of~\eqref{master} we thus have, for example $\delta J^1 g^{-1} J^1 = I^1\delta J^1$. Inserted back into \eqref{master} we arrive at
\begin{equation}
\delta g = - I^1 \delta J^1 - I^2 \delta J^2 - I^3 \delta J^3  = - I^x \delta J^x \ , 
\label{deltagf}
\end{equation}
or restoring the indices, $\delta g_{mn} = - {(I^x)_m}^p (\delta J^x)_{pn}$.

We can now apply eq.~\eqref{deltagf} to the computation of the line element in the space of metric deformations
\begin{equation}
\begin{aligned}
\delta s^2 & = \int_{Y_4} \! \sqrt{\vert g \vert} g^{mn} g^{pq} \delta g_{mp} \delta g_{nq} = \int_{Y_4} \! \sqrt{\vert g \vert} \, \tr (g^{-1} \delta g g^{-1} \delta g) \\ 
& = \int_{Y_4} \! \sqrt{\vert g \vert} \, \tr (g^{-1} I^x \delta J^x g^{-1} I^y \delta J^y) = \int_{Y_4} \! \sqrt{\vert g \vert} \, \tr (g^{-1} \delta J^x g^{-1} \delta J^x) \\ 
& = 2 \int_{Y_4} \delta J^x \wedge \ast \delta J^x \ .
\end{aligned}
\label{linee}
\end{equation}
In the last equations we used again $I^x \delta J^y g^{-1} = \delta J^y g^{-1} I^x$.

The next step is to express the (independent) variations $\delta J^x$ in terms of variations of the moduli $\delta \xi^x_A$. In particular we need to take into account the fact that variations which simply rotate the $J^x$ into themselves are physically equivalent. For such variations we must certainly have $\delta g_{mn} = 0$. We therefore require that the `physical' variations $\delta \xi^x_A$ are orthogonal to the $\xi^x_A$ or in other words they have to satisfy $\eta^{AB} \xi^x_A \delta \xi^y_B = 0$. (Note that these variations automatically respect the constraint~\eqref{orthoxi} and give us precisely the 9 restrictions that reduce the number of moduli contained in the 66 parameters $\xi^x_A$ to 57.\footnote{If the number of forms $\omega^A$ is kept general as $n + 6$ then the number of moduli is $3 (n + 3)$ which is the dimension of the Grassmanian~\eqref{su2coset}.}) The operator which projects onto this orthogonal subspace is given by $(\delta^A_B - \tfrac12 \xi^{yA} \xi^{yB})$.  Therefore the physically inequivalent variations of $J^x$ (apart from the variation of the volume) can be written as
\begin{equation}
\delta J^x = \e^{-\frac\rho2} (\delta^A_B - \tfrac12 \xi^{yA} \xi^{yB}) \delta \xi^x_B \omega^B \ , 
\label{varJ2}
\end{equation}
with $\xi^{xA} = \eta^{AB} \xi^x_B$ and $\delta \xi^x_B$ being unrestricted. Now we substitute eq.~\eqref{varJ2} into~\eqref{linee} and use~\eqref{Msym} and \eqref{M} to obtain
\begin{equation}
\delta s^2 = - 2 \e^{-\rho} (\eta^{AB} - \tfrac12 \xi^{yA} \xi^{yB}) \delta \xi^x_A \delta \xi^x_B \ . 
\end{equation}

Finally an overall rescaling of the $J^x$ parameterized by $\delta\rho$ and given by $\delta J^x = - \frac12 \delta\rho J^x$ leads to $\delta g = - \frac12 \delta \rho g$. It is not difficult to see that if we include also this contribution we have
\begin{equation}
\delta s^2 = \e^{-\rho} (\delta \rho)^2 - 2 \e^{-\rho} (\eta^{AB} - \tfrac12 \xi^{yA} \xi^{yB}) \delta \xi^x_A \delta \xi^x_B \ . 
\label{metricf1}
\end{equation}
Making use of eq.~\eqref{M} we can rewrite the last result in terms of $\delta {M^A}_B$ as follows
\begin{equation}
\delta s^2 = \e^{-\rho} (\delta \rho)^2 - \tfrac12 \e^{-\rho} \delta{M^A}_B \delta {M^B}_A \ .
\label{metricf2}
\end{equation}

\section{Computation of Killing prepotentials $\mathcal{P}^x_I$}
\label{Pfacts} 

Although the Killing prepotential $\mathcal{P}^x_I$ does not directly contribute to the potential due to the vanishing of the last term in \eqref{Vsugra}, we can nevertheless compute it from its definition. For completeness we devote this appendix to the computation of $\mathcal{P}^x_I$ following the procedure given in Appendix~D of ref.~\cite{Louis:2001uy}.

First of all lets introduce a $4\times24$ matrix $Z$ defined as
\begin{equation}
Z = \tfrac1{\sqrt2} \begin{pmatrix} \phantom{\Big\vert} \e^{\tfrac\rho2} & -\e^{-\tfrac\rho2} + \tfrac12 \e^{\tfrac\rho2} b^2 & - \e^{\tfrac\rho2} b^A \\ \phantom{\Big\vert} 0 & - \xi^x_A b^A & \xi^{xA} \end{pmatrix} 
\end{equation}
and satisfying $2 Z^\mathrm{T} Z = \mathcal{ML} + \mathcal{L}$ with $\mathcal{ML}$ and $\mathcal{L}$ given in eqs~\eqref{ML} and~\eqref{LL}, respectively. As second step lets define a $4\times4$ matrix $\Theta$ of one-forms as
\begin{equation}
\Theta = Z \mathcal{L}^{-1} \d Z^\mathrm{T} = \tfrac12 \begin{pmatrix} \phantom{\Big\vert} 0 & \e^{\tfrac\rho2} \xi^y_A \d b^A \phantom{\Big\vert} \\ \phantom{\Big\vert} - \e^{\tfrac\rho2} \xi^x_A \d b^A & \xi^x_A \d \xi^{yA} \phantom{\Big\vert} \end{pmatrix} \\ , 
\end{equation}
from which the $\mathrm{SU}(2)$ connection
\begin{equation}
\omega^x = - \tfrac12 \mathrm{tr} (\Theta \Sigma^x)
\end{equation}
follows. The three $4\times4$ matrices $\Sigma^x$ are the self-dual 't Hooft matrices as given in \cite{Andrianopoli:1996cm}. Now we compute the field strength for this connection, i.e. the triplet of two-forms
\begin{equation}
K^x = \d \omega^x + \tfrac12 \epsilon^{xyz} \omega^y \wedge \omega^z \ ,
\end{equation}
and solve the equation satisfied by the prepotentials $\mathcal{P}^x_I$, namely 
\begin{equation}
k^u_I K^x_{uv} = - (\partial_v \mathcal{P}^x_I + \epsilon^{xyz} \omega^y_v \mathcal{P}^z_I) \ . 
\label{momeq}
\end{equation}
It is not difficult to check that eq.~\eqref{momeq}, taking the Killing vectors $k^u_I$ for the general case as eqs.~\eqref{Killingv0} and~\eqref{Killingv} combined, has a solution given by 
\begin{equation}
\begin{aligned}
\mathcal{P}^x_{V^i} & = \tfrac12 (-1)^{x+1} \Big( \e^{\frac\rho2} b_A T^A_{iB} \xi^{xB} - \tfrac12 \epsilon^{xyz} \xi^y_A T^A_{iB} \xi^{zB} \Big) \ , 
\end{aligned}
\label{prepothyp}
\end{equation}
for a general $T_i$ of the form~\eqref{Ti}. This result can also be written as the integral expression
\begin{equation}
\mathcal{P}^x_{V^i} = \tfrac12 (-1)^x \epsilon_{ij} \e^\rho \Big[ \int_Y \d B \wedge J^x \wedge v^j - \tfrac12 \epsilon^{xyz} \int_Y J^y \wedge \d J^z \wedge v^j \Big] \ . 
\label{prepot}
\end{equation}
It can be checked that the Killing prepotential~\eqref{prepothyp} actually satisfies $\mathcal{P}^x_I = k^u_I \omega^x_u$.



\end{document}